\documentclass[
    ,final            
  ]
  {aipproc}

\layoutstyle{6x9}

\begin{document}

\title{Progenitors of type Ia supernovae in elliptical galaxies}


\author{M.Gilfanov}{
  address={Max-Planck-Institut f\"ur Astrophysik,  Garching bei M\"unchen, Germany}
 ,altaddress={Space Research Institute, Russian Academy of Sciences,  Moscow, Russia}
}

\author{\'A. Bogd\'an}{
  address={Harvard-Smithsonian Center for Astrophysics, Cambridge, MA, USA}
  ,altaddress={Max-Planck-Institut f\"ur Astrophysik,  Garching bei M\"unchen, Germany}
}

\begin{abstract}
Although there is a nearly universal agreement that type Ia supernovae are associated with the thermonuclear disruption of a CO white dwarf, the exact nature of their progenitors is still unknown. The single degenerate scenario envisages a white dwarf accreting matter from a non-degenerate companion in a binary system. Nuclear  energy of the accreted matter is released in the form of electromagnetic radiation or gives rise  to numerous classical nova explosions {\em prior} to the supernova event. We show that combined X-ray output of supernova progenitors and statistics of classical novae predicted in the single degenerate scenario are inconsistent with X-ray and optical observations of nearby early type galaxies and galaxy bulges. White dwarfs accreting from a donor star in a binary system and detonating at the Chandrasekhar mass limit can account for no more than $\sim 5\%$  of type Ia supernovae observed in old stellar populations.
\end{abstract}

\maketitle


\begin{figure}
\hbox{
  \includegraphics[height=.35\textheight]{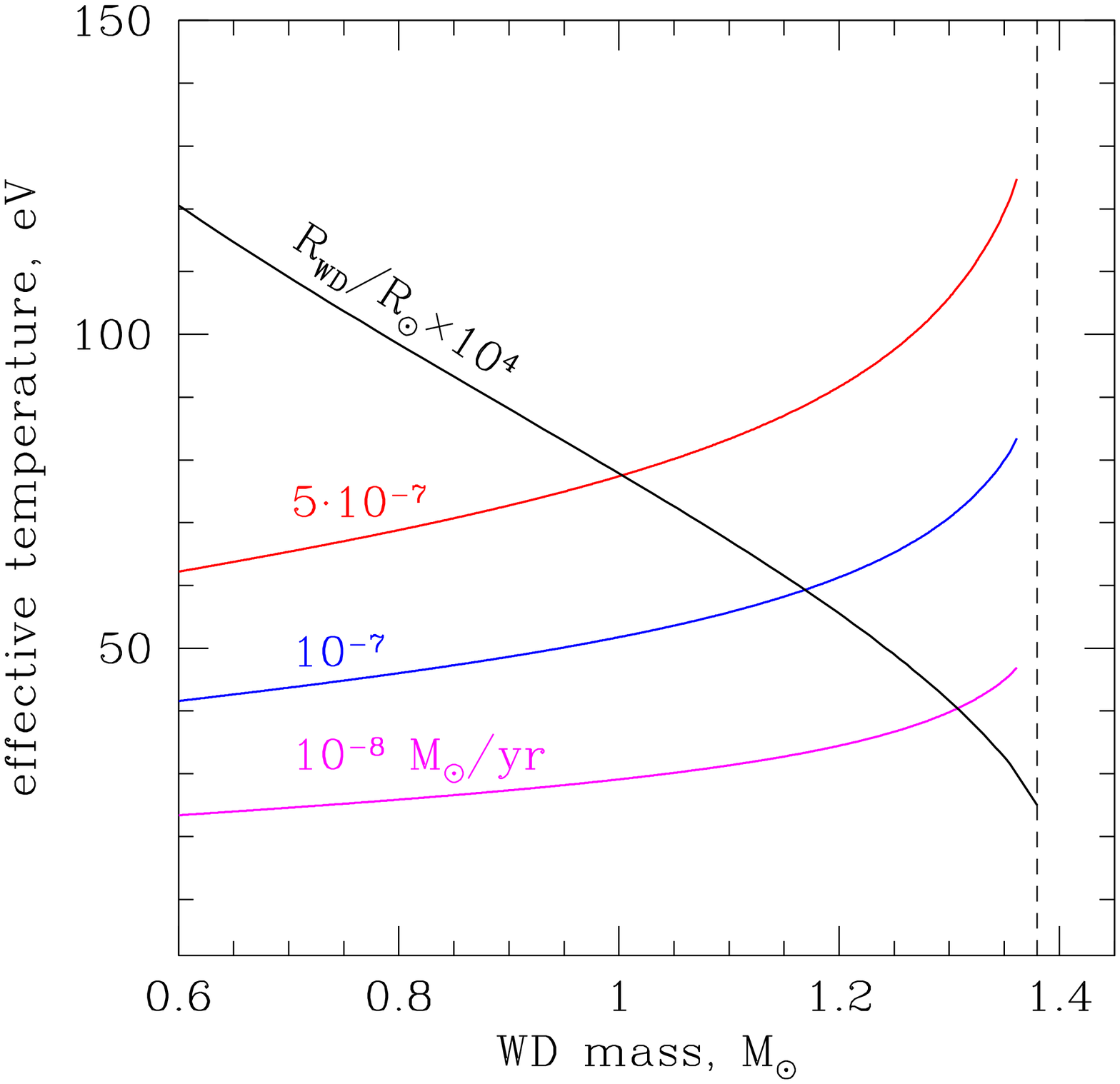}
  \includegraphics[height=.35\textheight]{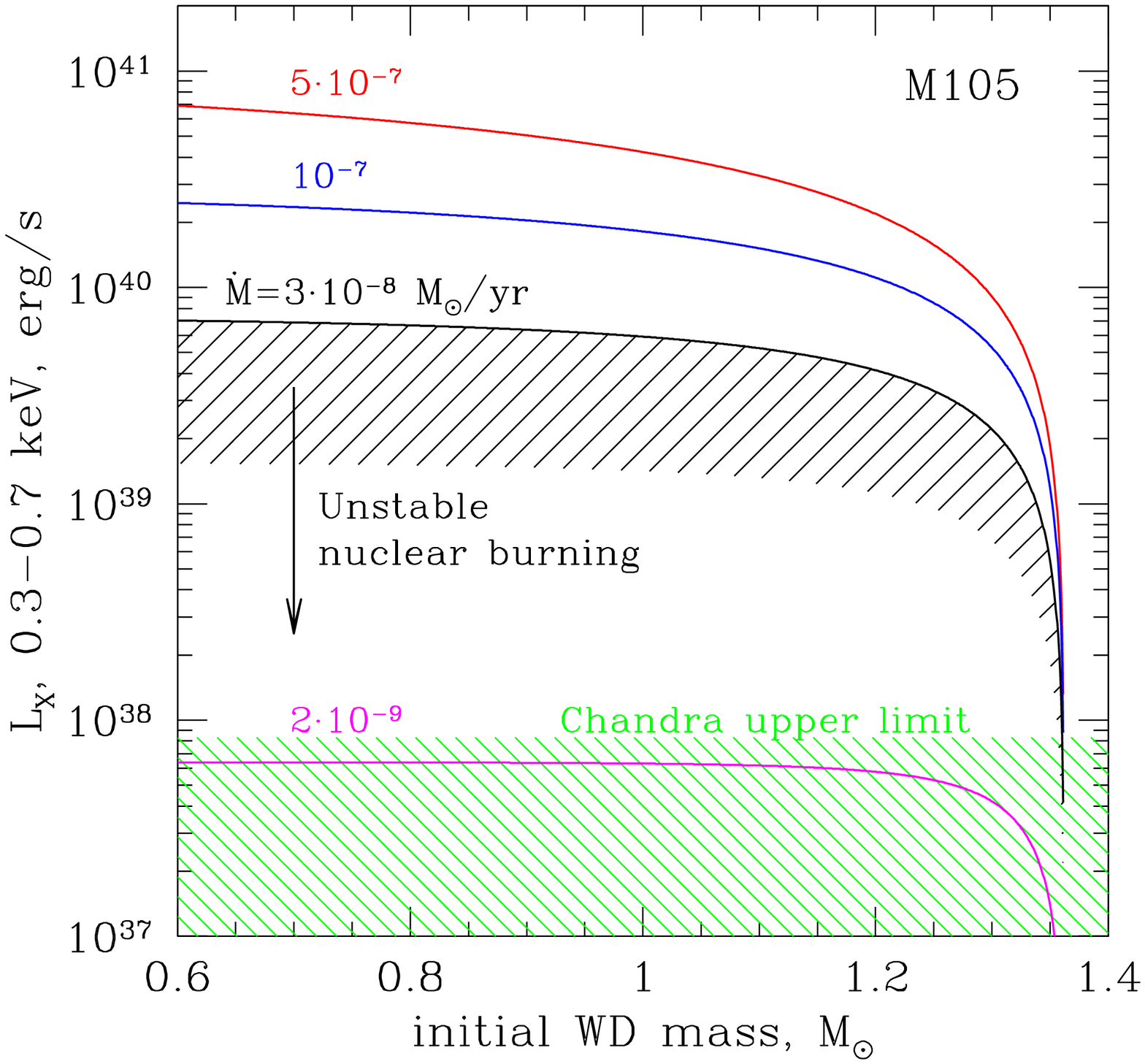}
  }
  \caption{{\em Left:} Dependence of the WD radius \cite{panei} and effective temperature of its surface on the WD mass. The effective temperature is computed for 3 different values of the mass accretion rate, assuming steady nuclear burning. {\em Right:} The expected soft X-ray luminosity of nuclear burning white dwarfs  in M105 galaxy assuming that all type Ia supernovae are formed according to the  accretion scenario.  The solid curves show predicted luminosity as a function of the initial WD mass, assuming that it is same for all SN Ia progenitors and they all have same $\dot{M}$ indicated by the numbers near the curves. 
The shaded area at the bottom shows the luminosity range compatible with Chandra observations. Its upper bound is an absolute upper limit which includes unresolved emission and combined luminosity of all sources with the color temperature $kT_c \le 200$ eV \cite{bogdan2010}. The actual luminosity of accreting WDs is likely by a factor of $\sim 2-3$ lower. In the single degenerate scenario plausible SN Ia progenitors have $\dot{M}\ge 10^{-7}	M_\odot/$yr \cite{livio_rev}.  Below $\sim{\rm few}\cdot 10^{-8}M_\odot/$yr CN outbursts will be triggered, which frequency will  contradict to observations (Fig.\ref{fig:cn}). 
\label{fig:kt}
\label{fig:lx} 
}
\end{figure}

A carbon-oxygen white dwarf (WD) formed through standard stellar evolution can not be more massive than  $\approx 1.1-1.2M_\odot$ \cite{weidemann}. Sub-Chandrasekhar models have been unsuccessful so far in reproducing observed properties of type Ia supernovae (SNe Ia) \cite{hoeflich96,nugent97}, although the effort continues \cite{sim2010}. In order to reach the Chandrasekhar mass of $\approx 1.38 M_\odot$  at least $\Delta M\ge 0.2M_\odot$ of matter needs to be accreted. Accretion of hydrogen-rich material onto the white dwarf is accompanied by hydrogen fusion  on its surface,  which is known to be unstable at small values of the mass accretion rate,  giving rise to  Classical Nova events \cite{nomoto07}. Because most of the accreted envelope and some of the original WD material is likely to be lost lost in the nova explosion  \cite{prialnik}, it is believed that  the WD does not grow in mass if  nuclear burning is unstable. For this reason  steady burning  regime is strongly favored by the accretion scenario \cite[e.g.][]{livio_rev}.

\subsection{X-ray constraints}

In steady nuclear burning regime, corresponding to the mass accretion rates  $\dot{M}\ge 10^{-7}~M_\odot$/yr,  energy of hydrogen fusion is liberated  in the form of electromagnetic radiation, with luminosity of 
$$L_{nuc}=\epsilon_H X \dot{M} \sim 10^{37} {\rm ~ erg/s}$$ 
where $\epsilon_H\approx 6\cdot 10^{18}$ erg/g  is energy release per unit mass and $X$ -- hydrogen mass fraction (the solar value of $X=0.72$ is assumed). 
The nuclear luminosity  exceeds by more than an order of magnitude the gravitational energy of accretion, $L_{grav}=GM\dot{M}/R$, and maintains the effective temperature of the WD  surface at the level, defined by the Stefan-Boltzmann law:  
\begin{equation}
T_{eff}\approx 67 
\left(\frac{\dot{M}}{5\cdot 10^{-7} M_\odot/yr}\right)^{1/4}
\left(\frac{R_{WD}}{10^{-2}R_\odot}\right)^{-1/2} 
{\rm~eV} 
\end{equation}
Such a soft spectrum  is prone to absorption by  interstellar gas and dust, especially at smaller temperatures. Because  the WD radius $R_{WD}$ decreases with its mass \cite{panei}, the  $T_{eff}$ increases as the WD approaches the Chandrasekhar limit --  the signal, detectable at X-ray wavelengths, will be dominated  by the most massive WDs (Fig.\ref{fig:kt}) \cite{nat2010, distefano2010apj}. 

The type Ia supernova rate  scales with near-infrared luminosity of the host galaxy  and for E/S0 galaxies $\dot{N}_{SNIa}\approx 3.5\cdot 10^{-4}$ yr$^{-1}$ per $10^{10}~L_{K,\odot}$ \cite{mannucci}.
If the WD mass grows at a rate $\dot{M}$,  a population of
\begin{equation}
N_{WD}\sim \frac{\Delta M}{\dot{M}\left<\Delta t\right>}\sim \frac{\Delta M}{\dot{M}}\dot{N}_{SNIa}
\label{eq:nwd}
\end{equation}
accreting WDs is needed in order for one supernova to explode on average every $\left<\Delta t\right>=\dot{N}_{SNIa}^{-1}$ years. Thus, for a typical galaxy, the accretion scenario predicts a numerous population of accreting white dwarfs, $N_{WD}\sim {\rm few} \times (10^2 -  10^3)$, much more than numbers of soft X-ray sources actually  observed \cite{distefano2010an}.
Therefore, although the brightest and hottest of them may indeed reveal themselves as super-soft sources, the vast majority of SN Ia progenitors must remain unresolved or hidden from the observer, for example by   interstellar absorption, in order for the accretion scenario to work. 
 Their combined luminosity  is 
\begin{equation}
L_{tot}=L_{nuc}\times N_{WD}=\epsilon X\Delta M\dot{N}_{SNIa}
\label{eq:ltot}
\end{equation}
where $\Delta M$ is the difference between the Chandrasekhar mass and  the initial WD mass. Predicted luminosity  can be compared with observations, after absorption and bolometric corrections are accounted for.

\begin{table} 
\caption{Comparison of the accretion scenario with observations \cite{nat2010}.
X-ray luminosities refer to the soft (0.3--0.7 keV) band. The columns marked "predicted" display total number and combined X-ray luminosity (absorption applied) of accreting WDs in the galaxy predicted in case the single degenerate scenario would produce all SNe Ia. In computing predicted numbers we halved the SN Ia rates as discussed in \cite{nat2010}, other parameters are: $\dot{M}=10^{-7}~M_\odot$/yr, initial WD mass $1.2M_\odot$. }
\vspace{0.5cm}
\centering
\begin{tabular}{l c c c c}
\hline
Name~~~~~~~~~  	& $L_K$ [$L_{K,\odot}$] & ~~~$N_{WD}$~~~ &\multicolumn{2}{c}{$L_X$ [erg/s]} \\
& ~~observed~~ & ~~predicted~~ &~~observed~~ & ~~predicted~~ \\
\hline
M32  &   $8.5\cdot 10^{  8}$ &  $25$ &$1.5\cdot 10^{36}$ & $7.1\cdot 10^{37}$	\\ 
NGC3377  &   $2.0\cdot 10^{10}$ &  $5.8\cdot 10^2$  &$4.7\cdot 10^{37}$ & $2.7\cdot 10^{39}$ \\
M31 bulge   &   $3.7\cdot 10^{10}$ &  $1.1\cdot 10^3$ &$6.3\cdot 10^{37}$ & $2.3\cdot 10^{39}$	\\
M105   &   $4.1\cdot 10^{10}$ 	&  $1.2\cdot 10^3$&$8.3\cdot 10^{37}$ & $5.5\cdot 10^{39}$  \\
NGC4278  &   $5.5\cdot 10^{10}$ &  $1.6\cdot 10^3$ &$1.5\cdot 10^{38}$ & $7.6\cdot 10^{39}$ \\
NGC3585  &   $1.5\cdot 10^{11}$  &  $4.4\cdot 10^3$ &$3.8\cdot 10^{38}$ & $1.4\cdot 10^{40}$ \\
\hline
\end{tabular}
\label{tab:lx}
\end{table}

To this end  we collected archival data of X-ray (Chandra) and near-infrared (Spitzer and 2MASS) observations of several nearby gas-poor elliptical galaxies and for the bulge of M31 (\cite{bogdan2010,nat2010}, Table \ref{tab:lx}).  Using K-band measurements to predict the SN Ia rates, we computed combined X-ray luminosities of SN Ia progenitors expected in the accretion scenario and compared them with Chandra observations. 
Obviously, the observed values present upper limits on the luminosity of the hypothetical population of accreting WD, as there may be other types of X-ray sources contributing to the observed emission.
As is clear from the Table \ref{tab:lx}, predicted luminosities surpass observed ones by a factor of $\sim 30-50$, i.e. the accretion scenario is inconsistent with observations by a large margin.

\subsection{Statistics of  recurrent and classical novae}

Unstable nuclear burning at low $\dot{M}$ is not generally believed to allow accumulation of mass, sufficient to lead to supernova explosion. However,  just below the stable burning limit a considerable fraction of the envelope mass can be retained by the WD during the nova explosion \cite{prialnik}. 
This motivated some authors to propose  recurrent novae as supernova progenitors  \cite[e.g.][]{hachisu2001}. 
 We demonstrate below that such systems will overproduce nova explosions in galaxies.

\begin{figure} 
\centering
\hbox{
\includegraphics[height=.35\textheight]{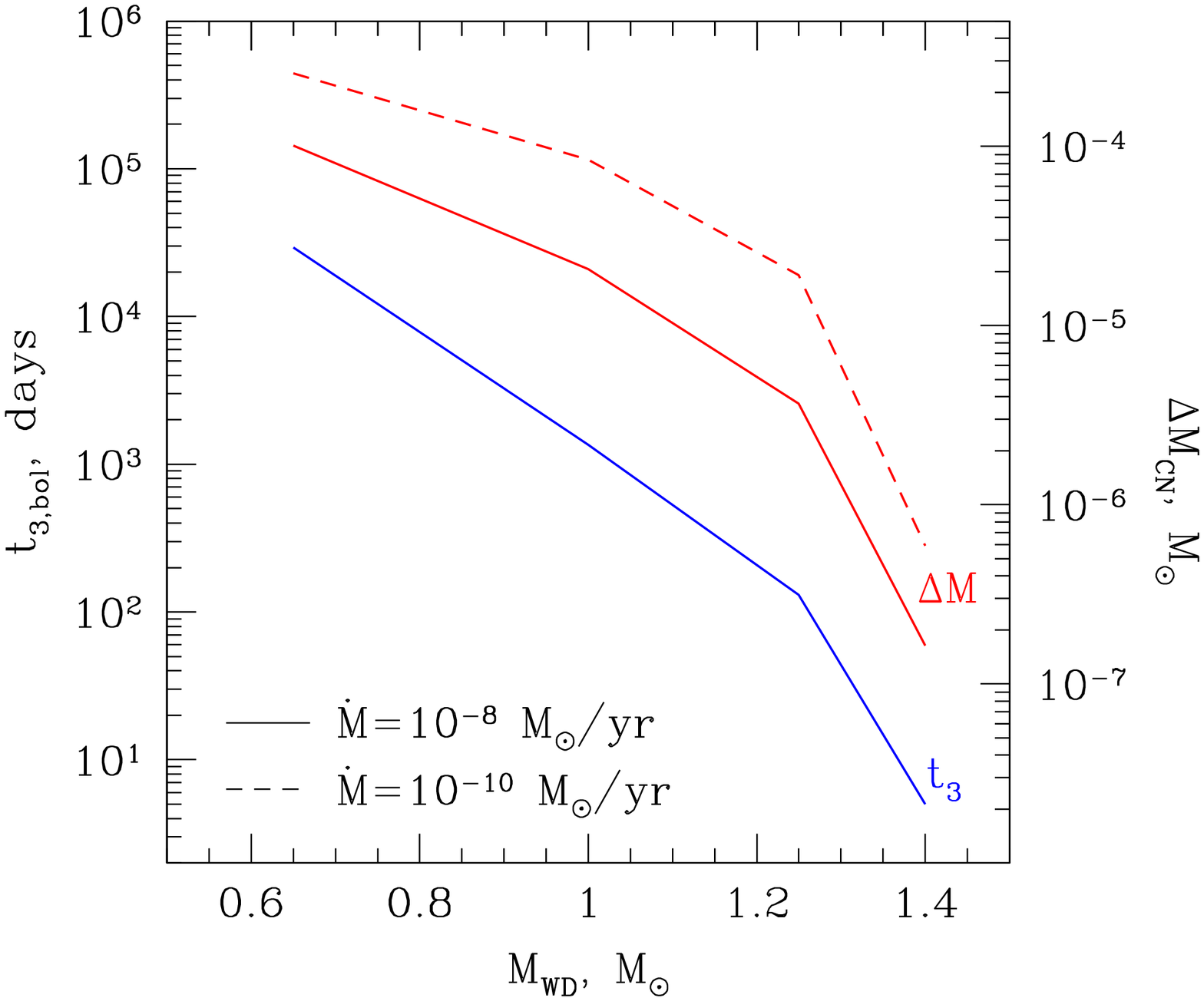}
\hspace{0.2cm}
\includegraphics[height=.30\textheight]{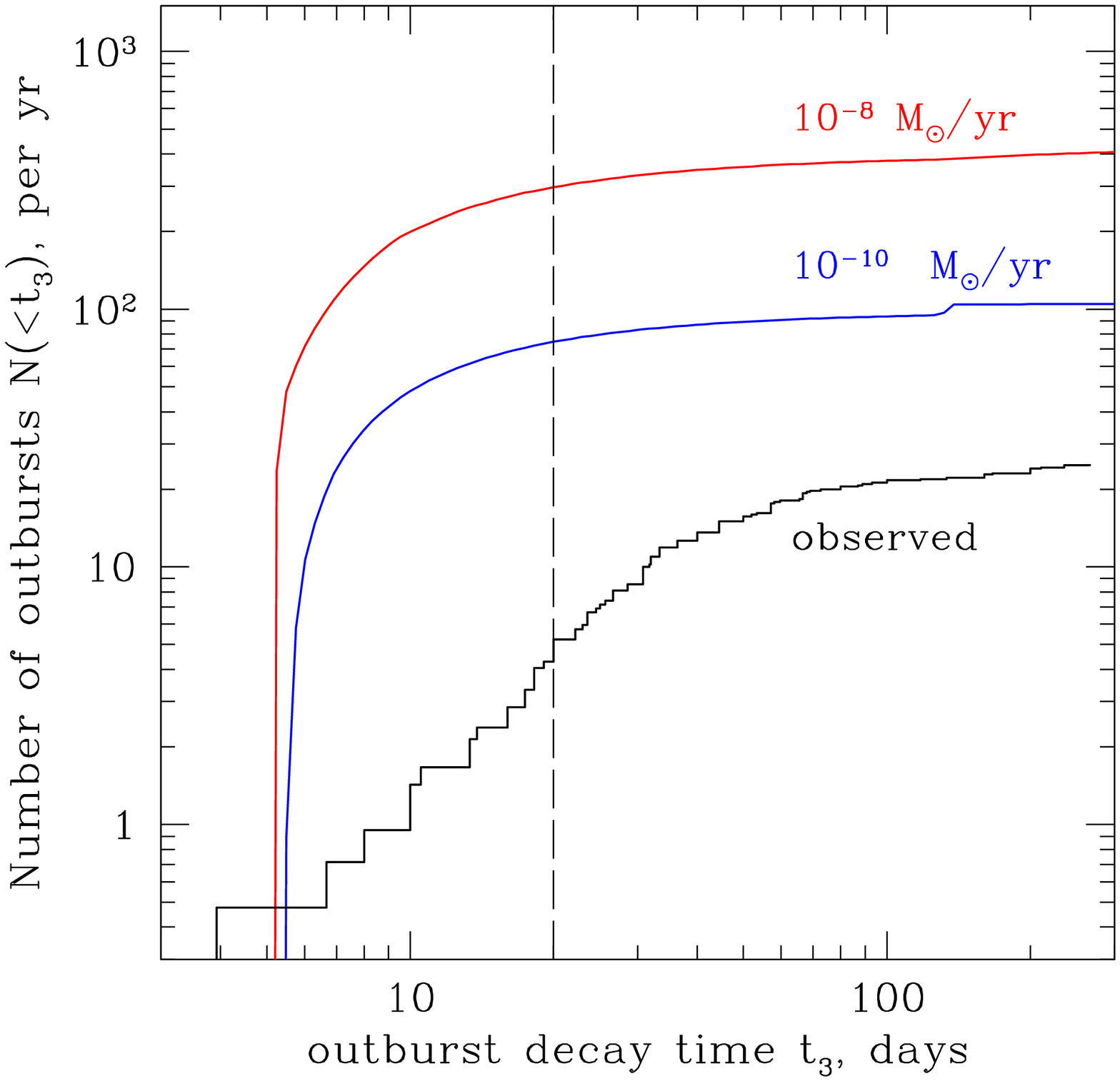}
}
\caption{{\em Left:} The $\dot{M}$ and $M_{WD}$ dependence of the nova ignition mass $\Delta M_{CN}$ and decay time
$t_3$, from \cite{prialnik}. {\em Right:} The decay time $t_3$ distribution for CNe in M31. The histogram shows the observed distribution  from \cite{capaccioli} normalized to the total rate of 25 CN/yr.  The systematic effects may somewhat modify its shape but will not affect our conclusions, as the observations control times do not vary significantly across the $t_3$ range of interest  \cite{capaccioli}.
Two smooth lines show predicted distributions for two different values of $\dot{M}$.  They are computed assuming that cataclysmic variables are the sole progenitors of SN Ia and are based on the results of numerical simulations of CN outburst cycles from  \cite{prialnik} for the WD core temperature of $10^7$ K. The vertical dashed line denotes the boundary of the fast Novae, which statistics is discussed in the text.  
}
\label{fig:cn}
\end{figure}

Assuming that classical/recurrent nova sources are the main type of SN Ia progenitors,
one can relate the nova and supernova rates:
\begin{equation}
\Delta M_{acc}\, \dot{N}_{CN} \sim \Delta M_{SN}\, \dot{N}_{SN}
\end{equation}
where $\Delta M_{acc}\le 10^{-7}-10^{-4}$ M$_\odot$ is the mass accumulated by the WD per one  nova outburst cycle,  
$\Delta M_{SN}\sim 0.5$ M$_\odot$ is the mass needed for the WD to reach the Chandrasekhar limit. 
As $\Delta M_{acc}$  depends on the $\dot{M}$ and WD mass (Fig.\ref{fig:cn}), we write more precisely:
\begin{equation}
\frac{\dot{N}_{CN}}{\dot{N}_{SNIa}}=\int\frac{dM_{WD}}{\Delta M_{acc}(M_{WD},\dot{M})}\ge
\int\frac{dM_{WD}}{\Delta M_{CN}(M_{WD},\dot{M})}
\label{eq:cn2sn}
\end{equation}
where $\Delta M_{CN}$ is the mass of the hydrogen shell required for the nova outburst to start. The inequality in the eq.(\ref{eq:cn2sn}) follows from the fact that $\Delta M_{acc}\le \Delta M_{CN}$, due to the possible mass loss during the nova outburst. 
As the $\Delta M_{CN}$ decreases steeply with the WD mass (Fig.\ref{fig:cn}), the main contribution to the predicted CN rate is made by the most massive WDs, similar to X-ray emission in the steady nuclear burning regime. They will be producing frequent outbursts with relatively short decay times \cite{prialnik}, thus resulting in a large population of fast (some of them recurrent) novae. This will contradict  to the statistics of CNe, as illustrated by the example of M31 shown in Fig.\ref{fig:cn}.  
Indeed, the observed rate of CN with decay time shorter than 20 days in this galaxy  is $\approx 5.2\pm1.1$ yr$^{-1}$ \cite{capaccioli},  while  eq.(\ref{eq:cn2sn}) predicts $\sim 300$ yr$^{-1}$ for  the mass accretion rates relevant to the recurrent novae-based progenitors models,  $\dot{M}\sim10^{-8}$ M$_\odot$/yr. As $\Delta M_{CN}$ is larger at small  $\dot{M}$ (Fig.\ref{fig:cn}), the contradiction between observed and predicted nova frequencies becomes less dramatic  at smaller $\dot{M}$.  However, very low values of $\dot{M}\ll 10^{-10}$ M$_\odot$/yr are  not feasible in the context of SN Ia progenitors. 
More realistic models with $\dot{M}\ge 10^{-8}$ M$_\odot$/yr do not produce more that $\sim 2\%$ of type Ia supernovae.

\subsection{Conclusion}

Thus, no more than $\sim$few per cent of SNe Ia in early type galaxies can be produced by\ white dwarfs accreting from a donor star in a binary system and exploding at the Chandrasekhar mass limit.
In the steady nuclear burning regime the supernova progenitors would emit too much of soft X-ray emission, while  if nuclear burning is unstable they would overproduce classical nova explosions.

At very high $\dot{M}$ the white dwarf could grow in mass without conflicting X-ray constraints or nova statistics, but would do this rather inefficiently, because a significant fraction of the transferred mass is lost in the wind \cite{hachisu, vdh97}. Therefore a relatively massive ({\em at least} $M\ge 1.3-1.7$ M$_\odot$) donor star is required for the white dwarf to reach the Chandrasekhar limit in this regime. Because the lifetimes of such stars do not exceed $\sim$few Gyr, this mechanism may work only in late-type and in the youngest of early-type galaxies.

As relevance of sub-Chandraskhar models  is still  debated \cite{hoeflich96, nugent97, sim2010}, the only currently  viable alternative are WDs mergers \cite{iben, webbink}. This mechanism may be the main formation channel for SN Ia in early type galaxies.  In late-type galaxies, on the contrary, massive donor stars are available, making the mass budget  less prohibitive, so that  WDs can grow to the Chandrasekhar mass entirely  inside  an optically thick wind or, via accretion of He-rich material from a He donor star \cite{iben94}. In addition, a star-forming environment is usually characterized by large amounts of neutral gas and dust, leading to increased absorption  obscuring  soft X-ray radiation from accreting WDs. 
Therefore in late-type galaxies  the accretion scenario may play a significant role, explaining, for example, the population of prompt supernovae \cite{bogdan2010a}.



\bibliographystyle{aipproc}   

\end{document}